\documentclass[12pt]{article}
\usepackage{amsmath}
\usepackage{amssymb}
\newenvironment{Abstract}
  {\begin{center}\textbf{Abstract}%
   \end{center} \begin{quote}\footnotesize}
  {\end{quote}}

\newcommand{\Imag}{\mathop{\rm Im}\nolimits}
\newcommand{\diver}{\mathop{\rm div\!}\nolimits}
\def\pderiv#1#2{\frac{\partial #1}{\partial #2}}
\def\aderiv#1#2{\frac{d #1}{d #2}}

\begin{document}
{\Large \centerline{Small signal gain analysis}
 \centerline{ for a wiggler with noncollinear}
\centerline{laser wave and electron beam}
 }

\bigskip
\bigskip
\centerline{A.I.~Artemiev$^{a,b}$, D.N.~Klochkov$^{a,b}$,
G.~Kurizki$^b$, N.P.~Poluektov$^a$, N.Yu.~Shubin$^{a,c}$}
 \centerline{\small\it $^a$General Physics Institute RAS, 38 Vavilov street, Moscow, 119991 Russia}
 \centerline{\small\it $^b$Chemical Physics Department, Weizmann Institute of Science, Rehovot 76100, Israel}
 \centerline{\small\it $^c$Institute of Microprocessor Computer Systems RAS, 36/1 Nakhimovsky prospect,
 Moscow 117997 Russia}
\bigskip
\bigskip
\begin{Abstract}
The collective and single-electron amplification regimes of a
non-collinear free electron laser are analyzed within the
framework of dispersion equations. The small-signal gain and the
conditions for self-amplified excitations are found. The
collective excitations in a free electron laser are shown to be
favored by the non-collinear arrangement of the relativistic
electrons and the laser wave. Implications for free-electron
lasing without inversion are discussed.
\end{Abstract}

\section{Introduction}

In a free-electron laser (FEL)~\cite{Madey},~\cite{Brau}, the
accelerated motion of electrons in the pondermotive potential of
the combined field of the wiggler and the amplified
electromagnetic wave produces coherent stimulated radiation. The
influence of the pondermotive potential induces structuring of the
spatial density of electrons (''bunching'') on the scale of the
laser wavelength. As a result net emission is enhanced. The
electron beam in FELs is usually aligned along the amplified
electromagnetic wave and the wiggler. The {\em non-collinear
geometry} of the electron beam and laser waves structures the
spacial and momentum distributions of electrons in a way that
gives rise to a new amplification mechanism called free-electron
laser without inversion (FELWI)~\cite{Kurizki}-\cite{Rostov}.
These FELs use the advanced phase control to enhance the gain via
interference of radiation produced in the two wigglers. To extend
the analysis of FELWIs from the single-electron (Thompson) regime
to the collective (Raman) regime of electromagnetic excitation a
detailed study should be made of the amplification in a single
wiggler for a non-collinear arrangement of the electron beam and
the amplified wave. The goal of the current paper is to perform
this analysis. Investigation of the regimes and conditions for the
amplifications in a non-collinear FEL geometry is also interesting
from academic point of view. This investigation generalizes the
results of the book \cite{Rukhadze} and allows one to obtain the
applicability limits of a single-electron approach used in papers
\cite{Kurizki}-\cite{Rostov}. It can have implications for other
types of free-electron lasers as well.

An FELWI is composed of two magnetic wigglers being spatially
separated by a drift region with magnetic field. There is a small
angle between the axes of wigglers. The laser wave propagates in
the direction having angles $\theta_1$ and $\theta_2$ with the
axes of the first and the second wigglers, respectively. The
electron beam is directed at the angles $\alpha_1$ and $\alpha_2$
to the axes of the wigglers. In the drift region the electrons are
turned by the magnetic field. In this device the electrons having
different acceleration in the first wiggler enter the drift region
at different directions. A magneto-optics set then separates these
electrons and introduces delays in their entrance phases (times of
entrance) for the second wiggler. It was
shown~\cite{Kurizki}-\cite{Rostov} that one can control the time
of electron entrance to the second wiggler so that the gain $G$ of
FELWI as a function of the detuning from the resonance condition
$\Omega=\omega(v_0-v_{res})/c$ is mostly positive and thus $\int
G(\Omega)d\Omega>0$. These results were obtained using the
single-electron approximation (Thompson regime): the propagation
of a single electron through the FEL system was considered and the
resulting gain was averaged over the electron distribution. But it
is known that the change of the system geometry may influence the
type of excitation regime, i.e. it can lead to a change from the
single-electron amplification regime to the collective one and
vice versa. For example, in the paper \cite{Rostov1} the
collective regime was considered in a non-collinear wiggler filled
with an {\em overdense} homogeneous plasma.

\section{Basic equations}

In order to find an analytical solution of the problem we assume
without a loss of generality that all the electrons have equal
velocities at the entrance of the first wiggler. We use the
approach developed in plasma electronics \cite{Rukhadze}, which
naturally describes the beam oscillations.

Let us consider the induced radiation by an electron beam in the
first wiggler. We choose the coordinate system so that the axis
$0z$ coincides with the axis of the wiggler while the wiggler
vector-potential is parallel to the axis $0y$. By assuming that
the static magnetic field of a plane wiggler ${\bf A}_\mathsf{w}$
is independent of the transverse coordinates $x$ and $y$, we can
approximate it by harmonic function
\begin{equation}\label{eq:1}
{\bf A}_\mathsf{w}=A_\mathsf{w}{\bf e}_y,\quad\text{where}\quad
A_\mathsf{w}=A_0e^{-i{\bf k}_\mathsf{w}{\bf r}}+\text{c.c.},
\end{equation}
where ${\bf k}_\mathsf{w}=(0,0,k_\mathsf{w})$ is the wiggler wave
vector,``c.c.'' denotes complex conjugation, and ${\bf e}_y$ is a
unit vector along $y$ axis. The wiggler field causes the electron
oscillations along the $y$-axis, therefore such an electron
interacts most efficiently with a linearly polarized light wave.
So we assume that vector potential of laser wave has linear
polarization ${\bf A}_L=A_L(t,x,z){\bf e}_y$. In this case the
vector potential ${\bf A}_L$ defines the purely vortex part of the
field $\, \diver{\bf A}_L=0$, while the scalar potential
$\phi=\phi(t,x,z)$ defines longitudinal beam waves in the system.
The Maxwell equations can be written in the form:
\begin{align}\label{eq:2}
&\Delta_\parallel\phi\equiv(\partial_x^2+\partial_z^2)\phi=-4\pi\rho,
\\ &(c^2\partial_x^2+c^2\partial_z^2-\partial_t^2)A_L=-4\pi cj_y.
\label{eq:2a}\tag{\ref{eq:2}$'$}
\end{align}

The electron beam entering the wiggler is assumed to have uniform
density $n_b$ and uniform electron velocity
 ${\bf u}=(-u\sin\alpha;0;u\cos\alpha)$. Then the initial distribution
function can be written in the form
 $f_0=n_b\delta({\bf p}_0-m\gamma_0{\bf u})$. Here $e$ and $m$ are
the electron charge and mass, $\gamma$ is the Lorentz factor. The
integral over the initial coordinates with this initial
distribution function gives the charge and current densities for
beam with charge compensated
\begin{align}\label{eq:3}
&\rho=en_b\left\{\int\delta[x-x(t,x_0,z_0)]\delta[z-z(t,x_0,z_0)]
dx_0dz_0-1\right\}, \\ &j_y=en_b\int v_y(t,x_0,z_0)
\delta[x-x(t,x_0,z_0)]\delta[z-z(t,x_0,z_0)] dx_0dz_0.
\label{eq:3a}\tag{\ref{eq:3}$'$}
\end{align}
Here $x(t,x_0,z_0)$ and $z(t,x_0,z_0)$ are the solutions of
Hamilton equations
\begin{equation}\label{eq:4}
\Dot{\bf r}=\pderiv{H}{{\bf P}},\qquad\Dot{\bf P}=-\pderiv{H}{{\bf
r}}
\end{equation}
with initial conditions  ${\bf r}_\parallel(0)={\bf r}_{\parallel
0}$, ${\bf p}_\parallel(0)=m\gamma_0{\bf u}$.  ${\bf P}={\bf
p}+\frac ec{\bf A}$ is the canonical momentum;
$A=A_\mathsf{w}+A_L$ is a sum of vector potentials. The
Hamiltonian of the electron in the field
\begin{equation}\label{eq:5}
H=\sqrt{m^2c^4+c^2\left( {\bf P}-\frac ec{\bf A}\right)^2} +e\phi
=mc^2\gamma+e\phi
\end{equation}
does not depend on $y$: $\partial H/\partial y=0$, and so we
obtain the first integral
\begin{equation}\label{eq:6}
v_y=\left.-\frac{e}{mc}\frac{A(t,x,z)}{\gamma}
\right|_{\substack{x=x(t,x_0,z_0)
\\z=z(t,x_0,z_0)}}.
\end{equation}

We represent all vectors as sums of two components: the first
component (designated as ${\bf f}_\parallel=(f_x,0,f_z)$) being in
the plane  $xz$, and the second component (designated as $f_y{\bf
e}_y$) being parallel to the vector-potential or vector ${\bf
e}_y$ . The Hamilton equations~\eqref{eq:4} determine the electron
coordinate and velocity
\begin{align}\label{eq:7}
&\Dot{\bf r}_\parallel=\pderiv{H}{{\bf P}_\parallel}={\bf
v}_\parallel, \\ &\Dot{\bf v}_\parallel
=-\frac{e}{m\gamma}\left[\nabla_\parallel-\frac{1}{c^2}{\bf
v}_\parallel({\bf v}_\parallel\cdot\nabla_\parallel)\right]\phi-
\frac 12\left(\frac{e}{mc}\right)^2\frac{1}{\gamma^2}
\left[\nabla_\parallel+\frac{{\bf v}_\parallel}{c^2}\pderiv{}{t}
\right]A^2
 \;.
 \tag{\ref{eq:7}$'$}\label{eq:7a}
\end{align}

We introduce two relativistic factors
\begin{equation}\label{eq:8}
\gamma_\parallel=\left(1-\frac{v_\parallel^2}{c^2}\right)^{-1/2},
\quad\gamma=\gamma_\parallel\left[1+\frac{1}{c^2}\left(\frac{e}{mc}
\right)^2A^2\right]^{1/2}.
\end{equation}

The field equations \eqref{eq:2} take the form
\begin{align}
\begin{split}\label{eq:9} \Delta_\parallel\phi=-\frac me\omega^2_b&\left\{
\int\delta[{\bf r}_\parallel-{\bf r}_\parallel(t,{\bf
r}_{\parallel 0})]d{\bf r}_{\parallel 0}-1 \right\},
\end{split} \\
\begin{split}\label{eq:9a}
(c^2\Delta_\parallel-\partial_t^2)A_L-&\omega_b^2\int\frac{A_L}{\gamma}
\delta[{\bf r}_\parallel-{\bf r}_\parallel(t,{\bf r}_{\parallel
0})]d{\bf r}_{\parallel 0}= \\
=&\omega_b^2\int\frac{A_\mathsf{w}}{\gamma} \delta[{\bf
r}_\parallel-{\bf r}_\parallel(t,{\bf r}_{\parallel 0})]d{\bf
r}_{\parallel 0}.
\end{split}\tag{\ref{eq:9}$'$}
\end{align}
Here $\omega_b^2=4\pi e^2n_b/m$ is square of Langmuir frequency of
the electron beam; here and below $\gamma=\gamma(t,{\bf r}
_{\parallel 0})$.

We look for the solutions for the field in the forms
\begin{align}\label{eq:11}
 &\phi=\frac 12\left[\psi e^{i{\bf k}_0{\bf r}_\parallel}
 +\text{c.c.}\right], \\ &A_L=A_+e^{i({\bf k}_0-{\bf k}_\mathsf{w})
  {\bf r}_\parallel}+A_-e^{-i({\bf k}_0+{\bf k}_\mathsf{w}){\bf r}_\parallel}.
 \tag{\ref{eq:11}$'$}\label{eq:11a}
\end{align}
Here vector ${\bf k}_0=k_0(\sin\theta,0,\cos\theta)$ lies in plane
$xz$. We denote the dimensionless coordinates as $\xi={\bf
k}_0{\bf r}_\parallel$, $\xi_0={\bf k}_0{\bf r}_{\parallel 0}$ and
 introduce the
 dimensionless  spatial
 Fourier-components of the electron charge and current density,
 $\sigma$ and  $\hat \sigma$, respectively:
\begin{equation}\label{eq:12}
\sigma=\frac{1}{\pi}\int\limits_0^{2\pi}e^{-i\xi}d\xi_0, \qquad
\hat{\sigma}=\frac{1}{\pi}\int\limits_0^{2\pi}
\frac{e^{-i\xi}}{\gamma}d\xi_0,
\end{equation}
Note that the integration is performed over the laser wavelength.
Here and below $\xi=\xi(t,\xi_0)$. Substituting  the solutions
\eqref{eq:11} in Eqs.\eqref{eq:9} and averaging these equations
over wavelength
%the electrons on wavelength (integrating over wavelength),
we get
\begin{align}\label{eq:13}
&\phi=\frac 12\frac me \frac{\omega_b^2}{k_0^2}\left[\sigma
e^{i\xi}+\text{c.c.} \right],
\\ &\frac{d^2A_+}{dt^2}+\omega_+^2A_++\omega_b^2I_0A_-=-\frac
12\omega_b^2\hat{\sigma}A_0, \tag{\ref{eq:13}$'$}\label{eq:13a}\\
&\frac{d^2A_-}{dt^2}+\omega_-^2A_-+\omega_b^2I_0^*A_+=-\frac
12\omega_b^2\hat{\sigma}^*A_0.\tag{\ref{eq:13}$''$}
 \label{eq:13b}
\end{align}
where
\begin{equation}\label{eq:14}
\begin{split}
&\omega_{\pm}^2=({\bf k}_0\mp{\bf
k}_\mathsf{w})^2c^2+\omega_b^2\langle\gamma^{-1}\rangle, \\
&I_0=\frac{1}{2\pi}\int\limits_0^{2\pi}
\frac{e^{-2i\xi(t,\xi_0)}}{\gamma(t,\xi_0)} d\xi_0, \\
&\langle\gamma^{-1}\rangle=\frac{1}{2\pi}\int\limits_0^{2\pi}
\frac{d\xi_0}{\gamma(t,\xi_0)}.
\end{split}
\end{equation}
Eqs.~\eqref{eq:13a}, \eqref{eq:13b} are the equations of
stimulated oscillations of two coupled  systems: the electron beam
and the amplified electromagnetic field.

The equations of the electron motion take the form
\begin{align}
\begin{split}\label{eq:15} \Dot{\bf v}_\parallel&=-\frac
i2\frac{\omega_b^2}{k_0^2} \frac 1\gamma\left[{\bf
k}_0-\frac{1}{c^2}{\bf v}_\parallel({\bf k}_0 {\bf
v}_\parallel)\right]\sigma e^{i\xi}- \\
&-\left(\frac{e}{mc}\right)^2\frac{e^{i\xi}}{\gamma^2} \left(i{\bf
k}_0+\frac{{\bf v}_\parallel}{c^2}\aderiv{}{t}\right)
(A_0^*A_++A_0A_-^*)+\text{c.c.},
\end{split} \\
\begin{split}\label{eq:15a}
\Dot{\bf r}_\parallel={\bf v}_\parallel,
\end{split}\tag{\ref{eq:15}$''$}
\end{align}
with the initial conditions ${\bf r}_\parallel(t=0)={\bf
r}_{\parallel 0}$, ${\bf v}_\parallel(t=0)={\bf u}$. The
self-consistent system of Eqs.~\eqref{eq:12}--\eqref{eq:15}
determines the stimulated radiation in the wiggler and describes
both {\em linear} and {\em nonlinear regimes} of the FEL
instability.

\section{Small signal gain}

\subsection{Dispersion equation}

Further we consider the {\em linear stage of instability} (small
signal gain). We linearize Eqs.~\eqref{eq:12}-\eqref{eq:15} for
small perturbations $\delta{\bf r}$, $\delta{\bf v}$, which are
proportional to the amplitudes of the laser waves $A_\pm$.
 All values are expanded in sums of non-disturbed and disturbed
 components:
  ${\bf r}_\parallel={\bf r}_{\parallel 0}+{\bf u}t+\delta{\bf r}_\parallel$ or
$\xi=\xi_0+{\bf k}_0{\bf u}t+{\bf k}_0\delta{\bf r}_\parallel$,
${\bf v}_\parallel={\bf u}+\delta{\bf v}_\parallel$, $\omega={\bf
k}_0{\bf u}+\Delta_\omega$, $\gamma=\gamma_0+\delta\gamma$
 and
$\gamma_\parallel=\gamma_{\parallel 0}+\delta\gamma_\parallel$.
Here
\begin{equation}\label{eq:17}
\gamma_0=\gamma_{\parallel 0}\sqrt{1+\mu},\qquad \gamma_{\parallel
0} =\left(1-\beta^2\right)^{-1/2},
\end{equation}
where $\beta=u/c$. The wiggler parameter $\mu$, which will play a
significant role, is defined as dimensionless square of the
wiggler field amplitude
\begin{equation}\label{eq:18}
\mu=\frac 2{c^2}\left(\frac{e}{mc}\right)^2|A_0|^2
\end{equation}
By linearizing equations over small perturbations, we obtain
$I_0=0$
 and
\begin{alignat}{2}\label{eq:19}
\sigma&=\delta\sigma e^{-i{\bf k}_0{\bf u}t},\qquad
&\delta\sigma=\frac 1\pi\int\limits_0^{2\pi}(-i{\bf k}_0\delta{\bf
r}_\parallel)e^{-i\xi_0}d\xi_0 \\ \hat{\sigma}&=\delta\hat{\sigma}
e^{-i{\bf k}_0{\bf u}t}, \qquad &\delta\hat{\sigma}=
\frac{\delta\sigma}{\gamma_0}-\frac{1}{\pi\gamma_0}
\int\limits_0^{2\pi}\frac{\delta\gamma}{\gamma_0}e^{-i\xi_0}d\xi_0
\tag{\ref{eq:19}$'$}\label{eq:19a}
\end{alignat}

For the small signal gain the vector-potential is a harmonic
function of time
\begin{equation}\label{eq:20}
A_\pm=a_\pm e^{\mp i\omega t}.
\end{equation}
The frequency $\omega$ is complex and its imaginary part defines
the growth rate of the FEL instability.

The solution to the linearized equations of motion~\eqref{eq:15}
follows:
\begin{align}\label{eq:21}
&\delta{\bf v}_\parallel=\left(\frac{e}{mc}\right)^2
\frac{e^{i\xi_0}}{D_b\gamma_0^3} \left(\beta_1{\bf k}_0
-\frac{\omega}{c^2}\beta_2{\bf u}\right) (A_0^*a_++A_0a_-^*)
e^{-i\Delta_\omega t}+\text{c.c.}
 \\
  &\delta{\bf r}_\parallel =
  i\left(\frac{e}{mc}\right)^2
\frac{e^{i\xi_0}}{D_b\gamma_0^3\Delta_\omega} \left(\beta_1{\bf
k}_0-\frac{\omega}{c^2}\beta_2{\bf u}\right) (A_0^*a_++A_0a_-^*)
e^{-i\Delta_\omega t}+\text{c.c.}
\tag{\ref{eq:21}$'$}\label{eq:21a}
\end{align}
Here
\begin{equation}\label{eq:22}
D_b=(\omega-{\bf k}_0{\bf u})^2-\Omega_b^2
\end{equation}
is the dispersion function of electron beam wave associated with
the beam frequency $\Omega_b$, where
\begin{equation}\label{eq:23}
\Omega_b^2=\frac{\omega_b^2}{\gamma_0} \left[1-\frac{({\bf
k}_0{\bf u})^2}{k_0^2c^2}\right].
\end{equation}
The coefficients  $\beta_1$ and $\beta_2$ equal
\begin{equation}\label{eq:24}
\beta_1=\gamma_0(\omega-({\bf k}_0{\bf u}))-\frac{\omega_b^2 ({\bf
k}_0{\bf u})}{k_0^2c^2},\qquad \beta_2=\gamma_0(\omega-({\bf
k}_0{\bf u}))-\frac{\omega_b^2}{\omega}.
\end{equation}

The perturbations of the dimensionless charge density $\sigma$ and
the dimensionless current density $\hat{\sigma}$ follow:
 \begin{align}\label{eq:25}
 &\delta\sigma=2\left(\frac{e}{mc}\right)^2\frac{1}{D_b\gamma_0^2}
 \left(k_0^2-\frac{({\bf k}_0{\bf u})\omega}{c^2}\right)
 (A_0^*a_++A_0a_-^*)e^{-i\Delta_\omega t}, \\
 &\delta\hat{\sigma}=\frac{2}{c^2}\left(\frac{e}{mc}\right)^2
 \frac{c^2k_0^2-\omega^2+\omega_b^2\gamma_0^{-1}}
 {D_b\gamma_0\gamma_{\parallel 0}^2(1+\mu)}(A_0^*a_++A_0a_-^*)
 e^{-i\Delta_\omega t}. \tag{\ref{eq:25}$'$}\label{eq:25a}
\end{align}
Substituting \eqref{eq:25a} in field equation \eqref{eq:13a},
\eqref{eq:13b} we obtain the dispersion equation, which defines
relation $\omega=\omega({\bf k})$.

Let us consider the resonant case $\omega\approx\omega_+={\bf
k}_0{\bf u}-\Omega_b$, which corresponds to the maximal growth
rate of the FEL instability. In this case $A_-=a_-=0$. As a result
the dispersion equation takes the simple form
\begin{equation}\label{eq:26}
D_b(\omega^2-\omega_+^2)=\frac 12\omega_b^2\frac{\mu}{1+\mu}
\frac{c^2k_0^2-\omega^2+\omega_b^2\gamma_0^{-1}}{
\gamma_0\gamma_{\parallel 0}^2}.
\end{equation}
Here
\begin{equation}\label{eq:39}
\omega_+^2=({\bf k}_0-{\bf
k}_\mathsf{w})^2c^2+\frac{\omega_b^2}{\gamma_0}
\end{equation}
 The solution of the dispersion equation \eqref{eq:26} under the
 resonant condition gives the frequency $\omega$
\begin{equation}\label{eq:40}
\omega=\omega_++\delta\omega=({\bf k}_0{\bf u})-\Omega_b+
\delta\omega.
\end{equation}
The presence of the beam leads to the complex shift of frequency
$\delta\omega$ (where $|\delta\omega|\ll\omega_+$). Its imaginary
part is the growth rate.

For the resonant conditions described above, the dispersion
function of the beam and the detuning of the frequency from the
resonance are equal to $D_b=\delta\omega^2-2\delta\omega\Omega_b$,
$\Delta_\omega=\delta\omega-\Omega_b$, respectively.

We introduce the complex dimensionless shift frequency
$\delta=\delta\omega/\Omega_b$. The dispersional
equation~\eqref{eq:26} can be written in terms of $\delta$ as

\begin{equation}\label{eq:40add}
\delta^2(\delta-2)+\frac 12\frac{\mu}{1+\mu}
\frac{\omega_b^2}{\Omega_b^2\gamma_0\gamma_{\| 0}^2}\,\delta=|q|.
\end{equation}
Where
\begin{equation}\label{eq:44}
|q|=\frac 14\frac{\mu}{1+\mu} \frac{(1+\nu)^2}{\nu}
\left(\frac{k_0c}{{\bf k}_0{\bf u}}\right)^2
\frac{\omega_b^2}{\Omega_b^2\gamma_0\gamma_{\| 0}^2},
\end{equation}
and
\begin{equation}
\nu=\frac{\omega_b^2}{\omega_+\Omega_b\gamma_0}.
\end{equation}
 For a
non-relativistic beam ($\beta\ll 1$) the parameter $\nu$ reduces
to the ratio of the frequencies $\nu=\omega_b/\omega_+$, i.e. to
dimensionless Langmuir frequency. It is shown below that the
parameter $\nu$ defines the normal or anomalous behaviors of the
growth rate, while the parameter $|q|$ defines the regime of
instability (Raman or Thompson).

Note that for {\em collinear FEL geometry}, when
$\alpha=\theta=0$, and relativistic electron beams we get
$|q|\approx 0.25 \mu/(1+\mu)\cdot (1+\nu)^2/\nu$, that is the
parameter $|q|$ depends on $\gamma_{\| 0}$ only through the
intermediary value $\nu$. To the contrary, for {\em non-collinear
FEL geometry}, when $\alpha+\theta\neq 0$, and relativistic
electrons the parameter $|q|$ will strongly depend on $\gamma_{\|
0}$. For $\gamma_{\| 0}\sin(\alpha+\theta)\gg 1$ we obtain the
asymptotic
\begin{equation}\label{eq:42}
|q|\approx\frac{\mu}{1+\mu}\frac{(1+\nu)^2}{\nu}
\frac{1}{\gamma_{\| 0}^2\sin^2(2\alpha+2\theta)}.
\end{equation}

In addition to, for collinear FEL geometry with $\gamma_{\| 0}$
increasing the parameter $\nu$ grows as a function
$\nu=(\omega_b/\omega_+)\sqrt{\gamma_{\| 0}/(1+\mu)}\propto
\sqrt{\gamma_{\| 0}}$, while for non-collinear FEL geometry under
condition $\gamma_{\| 0}\sin(\alpha+\theta)\gg 1$ the parameter
$\nu$ drops as $\gamma_{\| 0}$ increasing, namely
$\nu=\omega_b/(\omega_+\sqrt{\gamma_ 0}\sin(\alpha+\theta))\propto
1/\sqrt{\gamma_{\| 0}}$. This distinction leads to different
dependance of the parameter $|q|$ on $\gamma_{\| 0}$: while for
collinear laser geometry we have $|q|\sim\sqrt{\gamma_{\| 0}}$
(for $\nu\ll 1$) and $|q|\sim 1/\sqrt{\gamma_{\| 0}}$ (for $\nu\gg
1$), then for non-collinear laser geometry under
ultra-relativistic conditions $\gamma_{\| 0}\sin(\alpha+\theta)\gg
1$ we have $|q|\sim\gamma_{\| 0}^{-3/2}$. As was shown later, this
mean that for relativistic electron beams ($\gamma_{\| 0}
\sqrt{\nu} \sin(\alpha+\theta)\gg 1$ for $\nu\ll 1$ and
$\gamma_{\| 0} \sin(\alpha+\theta)/ \sqrt{\nu}\gg 1$ for $\nu\gg
1$) propagating at a small angle to laser wave direction, {\em the
collective amplification is possible for any value of parameter
$\mu$} (as distinct from collinear wiggler geometry
\cite{Rukhadze}, for which Raman regime is absent for $\mu>1$),
that is for any lateral relativistic velocity of electrons.

Let us consider different regimes of excitation.

\subsection{Collective amplification}

For the {\em collective  regime}, when $|\delta\omega|\ll\Omega_b$
or $|\delta|\ll 1$, and for relativistic beam $\gamma_{\|
0}\sin(\alpha+\theta)\gg 1$ the dispersion equation
\eqref{eq:40add} reduces to the quadratic form
\begin{equation}\label{eq:41}
\delta^2-\frac 14\frac{\mu}{1+\mu} \frac{\delta}{\gamma_{\|
0}^2\sin^2(\alpha+\theta)}+\frac 12|q|=0
\end{equation}
leading to  the growth rate for the collective regime:
$\Imag(\delta)=\sqrt{|q|/2}$ or
\begin{equation}\label{eq:43}
\Imag(\delta\omega)=\frac
12\sqrt{\frac{\mu}{2(1+\mu)}}\frac{k_0c}{{\bf k}_0{\bf
u}}\frac{\sqrt{\Omega_b\omega_+}}{\gamma_{\| 0}}
\left(1+\frac{\omega_b^2}{\omega_+\Omega_b\gamma_0}\right).
\end{equation}

The condition for Raman (collective) amplification can be
rewritten as $|q|\ll 1$. Thus for {\em non-collinear FEL geometry}
under relativistic condition $\gamma_{\| 0}\sin(\alpha+\theta)\gg
1$ the collective regime holds for any lateral relativistic
velocity of electrons. The increasing of the longitudinal velocity
(or relativistic factor $\gamma_{\| 0}$) for the {\em
non-collinear FEL geometry} decreases the parameter $|q|$ and thus
leads to the collective regime of amplification, independently
from the value of the wiggler parameter $\mu$.

Consider asymptotic formulas for growth rates of the undulator
radiation in the case of ultra-relativistic electron beams,
$\gamma_{\| 0}\sin(\alpha+\theta)\gg 1$:
\begin{equation}\label{eq:add1}
\! \Imag(\delta\omega)=\left\{
\begin{array}{ll}
\frac 12\sqrt{\frac{\mu}{2(1+\mu)}}\frac{\sqrt{\omega_+\omega_b
\sin(\alpha+\theta)}}{\gamma_0^{1/4}\gamma_{\| 0}
 \cos(\alpha+\theta)}, &\frac{\mu}{1+\mu}\frac{1}
 {\gamma_{\| 0}^2\sin^2(\alpha+\theta)}\ll \nu\ll 1
  \\
\frac{1}{2\sqrt{2}}\frac{\sqrt{\mu}}{(1+\mu)^{7/8}}\frac{\omega_b^{3/2}}
{\gamma_{\| 0}^{7/4}\cos(\alpha+\theta)
\sqrt{\omega_+\sin(\alpha+\theta)}},
  &
  {  1\ll \nu\ll\frac{1+\mu}{\mu}
\gamma_{\| 0}^2\sin^2(\alpha+\theta)}.
\end{array}
\right. \!\!\!\!\!
\end{equation}
The first growth rate \eqref{eq:add1} is the usual one
\cite{Rukhadze} for collective regimes, since its dependence on
Langmuir beam frequency is $\omega_b^{1/2}$. The second growth
rate is described by dependence $\omega_b^{3/2}$. This anomalous
behavior is a result of energy phase equalizing, which takes place
both in collinear \cite{Rukhadze} and non-collinear wiggler
geometry. For a non-collinear FEL geometry the growth rate depends
on the geometric parameter $\sin(\alpha+\theta)$ yet. Note that
the condition $\nu\gg 1$ can hold for overdense ultra-relativistic
beam, when $(\omega_b/\omega_+)^2\gg
\sin(\alpha+\theta)/\sqrt{1+\mu}$. The condition of the
amplification with the second growth rate of Eq.\eqref{eq:add1}
can be written in the form
\begin{equation}\label{eq:add2}
\max\left\{1,\left( \frac{\mu}{(1+\mu)^{3/4}}
\frac{\omega_b}{\omega_+}\frac{1}{\sqrt{\sin(\alpha+\theta)}}
\right)^{\frac 25} \right\}\ll\gamma_{\| 0}\sin(\alpha+\theta)\ll
\frac{\omega_b}{\omega_+}\frac{(1+\mu)^{1/4}}{\sqrt{\sin(\alpha+\theta)}}.
\end{equation}
The formula~\eqref{eq:add1} shows that the increasing of
longitudinal velocity (or relativistic factor $\gamma_{\| 0}$) for
non-collinear wiggler geometry leads to excitation of collective
regime independently from the values $\mu$ and $\nu$.

\subsection{Single-electron amplification}

For the {\em single-electron amplification} (Thompson regime) the
frequency shift $|\delta\omega|$ is larger than beam frequency,
namely $|\delta\omega|\gg\Omega_b$ or $|\delta|\gg 1$, and the
dispersion equation \eqref{eq:40add} is cubic
\begin{equation}\label{eq:46}
\delta^3+\frac
12\frac{\mu}{1+\mu}\frac{\omega_b^2}{\Omega_b^2\gamma_0 \gamma_{\|
0}^2}\delta-|q|=0.
\end{equation}
The solution Eq.\eqref{eq:46}, being written for image part of
$\delta$, is
\begin{equation}\label{eq:50}
\Imag(\delta)=\frac{\sqrt{3}}{2}|q|^{1/3}.
\end{equation}
The above definition of Thompson type of amplification
($|\delta|\gg 1$) can be rewritten as $|q|\gg 1$.

Consider the asymptotic  of the growth rate  $\Imag(\delta\omega)$
for $\gamma_{\| 0}\sin(\alpha+\theta)\gg 1$, the case of interest
for FELWI applications. Under conditions
\begin{equation}\label{eq:add3}
\nu\ll\min\left\{\frac{\mu}{1+\mu}\frac{1}{\gamma_{\|
0}^2\sin^2(\alpha+\theta)}\, ,\, 1\right\}
\end{equation}
the asymptotic behavior of the growth  rate  is
\begin{equation}\label{eq:add4}
\Imag(\delta\omega)=\frac{\sqrt{3}}{2^{5/3}}\left[
\frac{\mu}{(1+\mu)^2}\frac{\omega_b^2\omega_+}{\gamma_{\| 0}^3}
\tan^2(\alpha+\theta)\right]^{1/3}.
\end{equation}
For very large  $\nu$, namely $\nu\gg \gamma_{\|
0}^2\sin^2(\alpha+\theta)(1+\mu)/\mu$, the growth rate of
single-electron amplification has the anomalous behavior
\begin{equation}\label{eqL:add5}
\Imag(\delta\omega)=\frac{\sqrt{3}}{2^{5/3}}
\frac{\mu^{1/3}}{\sqrt{1+\mu}}\left(\frac{\omega_b}{\omega_+}\right)^{1/3}
\frac{\omega_b}{{\gamma_{\| 0}^{4/3}} \cos^{2/3}(\alpha+\theta)}.
\end{equation}

As for {\em collinear FEL geometry}, here the growth rate depends
on Langmuir frequency of the electron beam as $\omega_b^{4/3}$ and
is almost independent from the angle between the electron beam and
the laser wave.

However, the realization of this amplification regime using
ultra-relativistic beam is almost impossible because of the large
required charge of beam $(\omega_b/\omega_+)\gg \gamma_{\|
0}^{5/2} \sin^3(\alpha+\theta)(1+\mu)^{3/4}/\mu$, and as a
consequence it is necessary very big current, which is limited by
{\em vacuum current} for vacuum devices.

The above calculations indicate that if the wiggler is loaded with
the non-collinear electron beam and the laser wave, then the
Raman-type amplification is feasible for relatively small
densities of the electron beam (the first growth rate in
Eq.\eqref{eq:add1}). We find that the electron current density
required for Raman-type amplification drops with increasing the
relativistic factor $\gamma_{\| 0}$ of beam. This means that {\em
collective amplification} can be realized in optical wigglers, in
particular, in FELWI, in which the ultra-relativistic
non-collinear beams are used.

\section{Conclusions}

Summarizing, we consider Thompson and Raman regimes of FEL
amplification for the non-collinear geometry of the electron and
laser beams. It was found that the non-collinear geometry shifts
the conditions for the amplification toward collective (Raman)
regime. It was found that if the wiggler is loaded with the
non-collinear electron beam and the laser wave, then the
Raman-type amplification is feasible for relatively small
densities of the electron beam (the first growth rate in
Eq.\eqref{eq:add1}). We find that the electron current density
required for Raman-type amplification drops with increasing the
relativistic factor $\gamma_{\| 0}$ of beam. This means that {\em
collective amplification} can be realized exactly in optical
wigglers, in particular, in an FELWI, which employs
ultra-relativistic electron beams non-collinear laser wave.

\section{Acknowledgements}

N.Yu.Sh., D.N.K. and A.I.A. gratefully acknowledge support by the
RFBR grant 02-02-17135 and support by the International Science
and Technology Center, Moscow, through the Project A-820.


{\small
\begin{thebibliography}{99}
\bibitem{Madey} J.M.J. Madey,
 Nuovo Cimento Soc. Ital. Fis. {\bf 50B}, 64 (1979).
\bibitem{Brau} C. A. Brau,
 {\em Free-Electron Lasers} (Academic, Boston, 1990).
\bibitem{Kurizki}G.~Kurizki, M.O.~Scully,
C.~Keitel, Phys.~Rev.~Lett. {\bf 70}, 1433 (1993).
\bibitem{Sherman}B.~Sherman, G.~Kurizki, Phys.~Rev.~Lett. {\bf75},
4602 (1995).
\bibitem{Nikonov1}D.E.~Nikonov, B.~Scherman, G.~Kurizki,
M.O.~Scully, Opt.~Commun. {\bf 123}, 363 (1996).
\bibitem{Nikonov2}D.E.~Nikonov, M.O.~Scully, G.~Kurizki,
Phys.~Rev.~E {\bf54}, 6780 (1996).
\bibitem{Nikonov3}D.E.~Nikonov, Yu.V.~Rostovtsev, G.~Sussmann,
Phys.~Rev.~E {\bf57}, 3444 (1998).
\bibitem{Rostov}Yu.~V.~Rostovtsev, S.~Trendafilov, A.~I.~Artemyev,
K.~T.~Kapale, G.~Kurizki, and M.~O.~Scully,  Phys.~Rev.~Lett. {\bf
90}, 214802 (2003).
\bibitem{Rostov1} A.B. Matsko, Yu.V. Rostovtsev, Phys. Rev. E {\bf
58}, 7846, (1998).
\bibitem{Rukhadze}  M.V. Kuzelev, A.A. Rukhadze
{\it Plasma Free Electron Lasers}. Edition Frontier, Paris, 1995.
\end{thebibliography}
}
\end{document}